\documentclass[aps,prb,twocolumn,groupedaddress,nofootinbib,10pt,longbibliography]{revtex4-1}
\usepackage{CJK}

\usepackage{hyperref}
\usepackage{xcolor}
\usepackage{braket}
\usepackage{slashed}

\usepackage{mathtools}
\usepackage{nicefrac}
\usepackage{amssymb}
\usepackage{mathrsfs}
\usepackage{IEEEtrantools}
\usepackage{tabularx}

\definecolor{BrickRed}{cmyk}{0, .89, .94, .28} 

\newcommand{\be}{\begin{equation}}
\newcommand{\ee}{\end{equation}}
\newcommand{\bea}{\begin{IEEEeqnarray}{rCl}}
\newcommand{\eea}{\end{IEEEeqnarray}}

\newcommand{\ba}{\begin{array}}
\newcommand{\ea}{\end{array}}

\begin{document}

\title{Modeling Resistive Switching in Nanogranular Metal Films}

\author{Walter Tarantino}
\email{walter.tarantino@dsf.unica.it}
\affiliation{Dipartimento di Fisica, Università degli Studi di Cagliari, Cittadella Universitaria, I-09042 Monserrato, Cagliari, Italy}
\author{Luciano Colombo}
\affiliation{Dipartimento di Fisica, Università degli Studi di Cagliari, Cittadella Universitaria, I-09042 Monserrato, Cagliari, Italy}

\date{\today}

\begin{abstract}

Films produced by assembling bare gold clusters
well beyond the electrical percolation threshold
show a resistive switching behavior whose investigation has started
only recently. 
Here we address the challenge to charaterize the resistance 
of a nanogranular film starting from limited information on the structure at the microscopic scale by the means of Bruggeman's approach to multicomponent media,
within the framework of Effective Medium Approximations.
The approach is used to build a model that proves 
that the observed resistive switching can be explained by
thermally regulated local structural rearrangements.
\end{abstract}

\maketitle
\section{Introduction}

The synthesis of materials and devices with new properties by means 
of controlled manipulation of their structure down to the atomic scale
is nowadays paving the way to a plethora of new technological applications.
Among these, neuromorphic computing architectures \cite{chen2018},
which are believed to have the potential of overcoming 
the looming end of Moore's scaling law,
are becoming possible thanks to `resistive switching junctions' \cite{lee2020},
that is,
devices that suddenly change their resistance under the action 
of an electric current in a non-volatile and reversible way.
Resistive switching (RS) behavior has been reported in 
several systems, from oxides to semiconductors
to organic materials \cite{lee2015,yang2013}.
In particular, networks of nanowires, nanoparticles and clusters 
can also be used to make RS devices. This can be achieved by creating
insulator-conductor nanocomposites, namely
by embedding the metallic components into a polymeric matrix
or passivating them by a shell of ligands or oxide layers \cite{lee2015,hwang2019}.
Purely metallic nanoparticle systems can also present RS behavior
if randomly deposited on a substrate to form a highly discontinuous film 
near the percolation threshold.\cite{borziak1976,sattar2013,minnai2017}
In all those systems, RS emerges as an insulator-to-metal transition
due to the formation of a conducting percolation network,\cite{lee2015,fostner2015}
which can be described in terms of percolation theory \cite{kirkpatrick1973,stauffer1994,sahimi1994}.

Recently, cluster-assembled gold films have been reported to present
an unexpected form of resistive switching \cite{mirigliano2019,mirigliano2020}.
By supersonic cluster beam deposition of bare Au nanoparticles
on various substrates,
metallic films characterized by a complex microstructure have been grown
well beyond their electrical percolation threshold.
Nanoparticles deposited on the substrate basically retain 
their individuality, thus forming films of (nano)granular matter (henceforth ``ng-films''),
a medium highly porous and rich of interfaces among misoriented crystal grains.
Besides increasing the electrical resistance of 
ordinary (i.e. continuous  or, equivalently, atom-assembled) thin films,
their nanostructure results in a remarkable dynamical
response to a sufficiently high applied voltage.
Even under the action of a direct electric current
a ng-film presents a resistance characterized by nearly ohmic regimes
(during which only small fluctuations occur)
and abrupt jumps to higher as well as to lower values,
after which the system either returns to the same ohmic regime (\emph{spikes})
or reaches a new one (\emph{steps}).
As the system is well above its electrical percolation threshold,
the underlying mechanisms differ from the usual insulator-to-metal transition
responsible for RS in other systems.
Therefore, they need to be elucidated by means of 
further experimental as well as theoretical
investigations and this work represents an attempt 
to build some appropriate theoretical
tools for such an endeavor.

The outcome of a complete theoretical characterization 
of such a RS phenomenon would be a model able to reproduce
the abrupt changes in resistance
and to provide a reasonable estimate of the spanned values.
The challenge 
is twofold.
On the one hand, one needs to identify and characterize 
the underlying mechanisms responsible for the jumps of resistance.
On the other hand, one must be able to 
estimate the resistance of a ng-film, which, 
even in the ohmic regime, is not at all an easy task to accomplish,
as the highly inhomogeneous nature of the system is not captured by
the standard models used to estimate the resistance 
of crystal films.
The aim of the present work is therefore to establish
a theoretical framework that allows to
estimate the resistance of a ng-film
starting from ingredients which could be accessible to 
\textit{ab initio} and semi-empirical methods.
In other words, the present attempt to elaborate
an adequate formalism linking the observed RS behavior
to some underlying microstructure evolution should
pave the way to a multi-physics approach, 
where the resulting theoretical model will be fed by physical 
information computed from first principles.

The first steps in that direction were made already 
in Ref. \onlinecite{mirigliano2019},
where the experimental results were qualitatively reproduced using 
a dynamical resistor network model.
The approach is very flexible and has been used 
to explain many emergent properties of nanoparticle and nanowire networks 
\cite{Diaz-Alvarez_2019,fostner2014,fostner2015,Loeffler_2020,Mallinson_2019,%
Manning_2018,Milano_2020,Pike_2020,sillin2013}. 
However, the computational cost of solving the Kirchhoff’s equations 
implied by a nanoscale resolved effective resistor network modeling 
a macroscopic three-dimensional sample can be very demanding. 
As we are interested in a regime far from the percolation threshold, 
we do not need to rely on such a precise description and we can resort to other, 
approximate and thus more affordable, methods. We therefore adopt
an Effective Medium Approximation (EMA),
which in principle allows to estimate the electrical resistivity
of a ng-film starting just from limited knowledge 
about its microscopic structure \cite{landauer1978,stroud1998,grimaldi2014}.
In Section \ref{sStat} we present the approach and explain how,
in a steady situation, it can be used
to account for the high degree of porosity of a structure, 
as well as for other structural features
that define the nanogranular character of the system. 
In Section \ref{sDyn} we add a dynamical evolution
due to 
thermally regulated structural changes,
and show that they indeed rule over the observed RS behavior; 
we also explain how they can be modeled within our framework.
Two realizations of this modeling
are shown to capture many features of the experimentally 
reported ng-film resistance \cite{mirigliano2019,mirigliano2020}.

Such an analysis allows us to trace a clear path for future investigations
aimed at identifying the actual atomic-scale mechanisms
occurring in the films.
Full extent and limits of the approach 
are discussed in the last section.


\section{Ohmic regime}\label{sStat}

\subsection{Synopsis of the residual resistivity contributions}

Before addressing the dynamical evolution of a ng-film,
we must discuss its conductance in a static regime.
Since its dynamical response is only obtained for high enough applied voltages,
the static regime can be taken as the ohmic regime occurring
when a weak induced current is used as a probe.
For stronger currents, the response is describes as
nearly ohmic regimes punctuated by jumps of resistance.
This, together with the observation that during the occurrence of the RS phenomenon
the film does not seem to undergo any qualitative structural 
change at the macroscopic scale,
allows us to hypothesize that the inhomogeneities characterizing the ohmic regime
at the microscopic scale remain qualitatively the same
also during the dynamical regime.
In other words, the sources of electronic scattering characterizing the ohmic regime
we are about to discuss remain the same during the dynamical one.
What changes is their quantitative contribution
to the overall resistance of the sample.

Taking apart the contribution of the temperature,
which, as we will see, we believe to play 
an important role in the dynamical evolution,
we focus for now on the residual resistivity (RR) of a ng-film.
It is known that for such systems the RR can be orders
of magnitude higher than that of films of the same metal 
in the crystalline phase \cite{hebard1999}.
In fact, the RR of these latter can already differ 
quite sensibly from that of the bulk material. 
Grain boundaries and surface effects are known to play a major role
\cite{gould2017} and the models of Mayadas-Shatzkes 
\cite{mayadas1970} and Fuchs-Sondheimer \cite{fuchs1938,sondheimer1952}
are routinely used to account for them.
In the case of ng-films, 
additional structural imperfections have a higher impact
on the overall RR.
We organize such imperfections in the following way.
Foremost we consider the contribution of voids,
namely the interstitial vacuum existing among nanoparticles.
Although they do not allow for band transport,
they still allow hopping and tunneling,
whose contribution to the overall current
is, however, expected to be much lower than band transport.
A second relevant source of RR
are the interfaces between nanoparticles.
In the best case, they can be assimilated to boundaries between
randomly oriented grains belonging to different nanoparticles.
Most likely, however, an amorphous layer is 
present in the contact region, as a result
of the impact during the deposition stage \cite{benetti2017}.
To best of our knowledge, 
there is no well-established method in literature to account
for neither imperfections in estimating the resistivity
of a film.

Surface effects, which may become relevant for thin films,
can be estimated using the Fuchs-Namba model \cite{namba1970},
an extension of Fuchs formalism \cite{fuchs1938} 
that also takes into account  the roughness 
of the top layer of nanoparticles.
The presence of grains, on the other hand,
is not adequately described by the model of Mayadas-Shatzkes,
which assumes that grains grow in a columnar structure, 
with the axes normal to the film plane, 
and extend from two surfaces of the film.
While usually valid for continuous crystal films,
these assumptions are not justified for
nanogranular ones, 
where grains are randomly oriented, as a result 
of the nanoparticle deposition, and therefore 
do not satisfy the assumptions of the model. 
Impurities and further defects inside the grains add the
last contribution to the RR.

\begin{figure}[ht]
\centering
\includegraphics[width=0.45\textwidth]{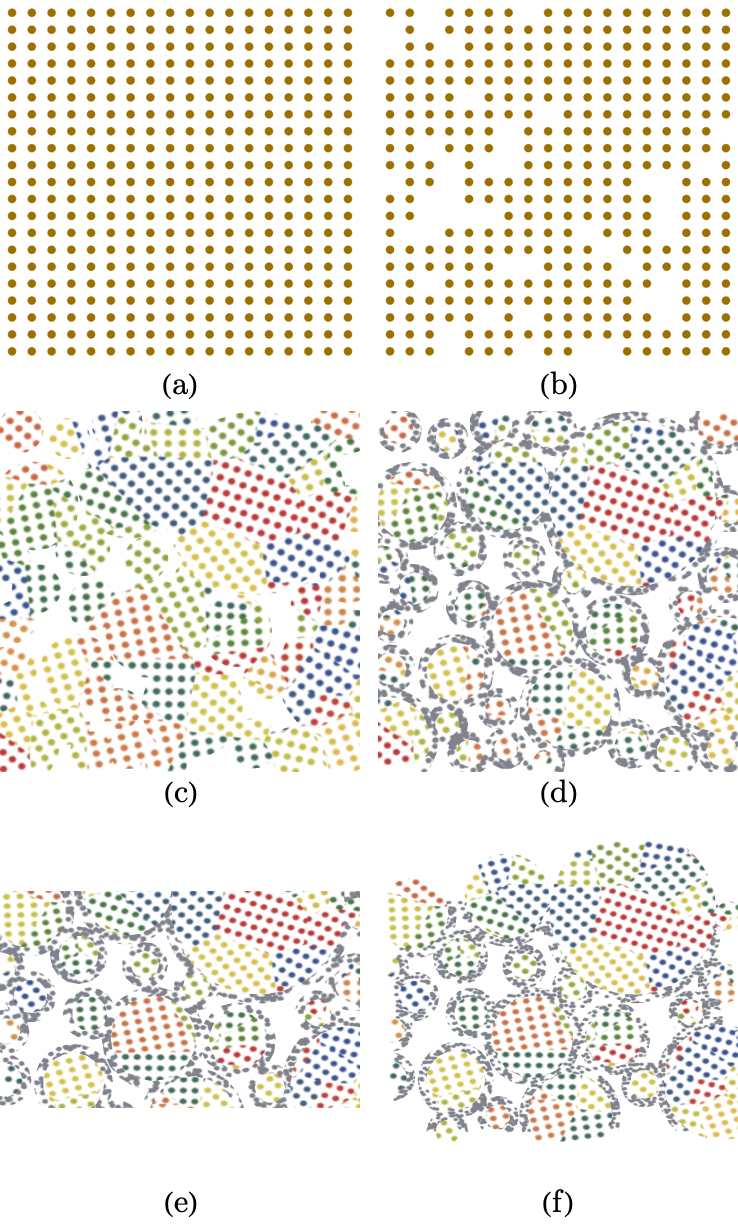}
\caption{
The RR of a ng-film can be regarded as arising 
from the contribution of a variety of sources 
of electronic scattering:
(a) the crystalline structure,
(b) the interstitial voids,
(c) the presence of randomly oriented grains 
(denoted by different colors),
(d) the presence of thick nanoparticle interfaces,
(e) surface effects due to the `thinness' of the film,
(f) roughness of the surface, impurities and other defects.
}\label{figMatthiessen}
\end{figure}

Such a hierarchy of contributions to the RR of 
a nanonagranular film is summarized in Fig. \ref{figMatthiessen}.
Since the RR determined by the imperfections pictorially depicted 
in (a), (e), and (f) can be accurately estimated by means 
of well-established theoretical tools, 
in the following we shall focus on the modeling of (b), (c), and (d).

\subsection{EMA approach}\label{sEMA}

\begin{figure}[ht]
\centering
\includegraphics[width=0.47\textwidth]{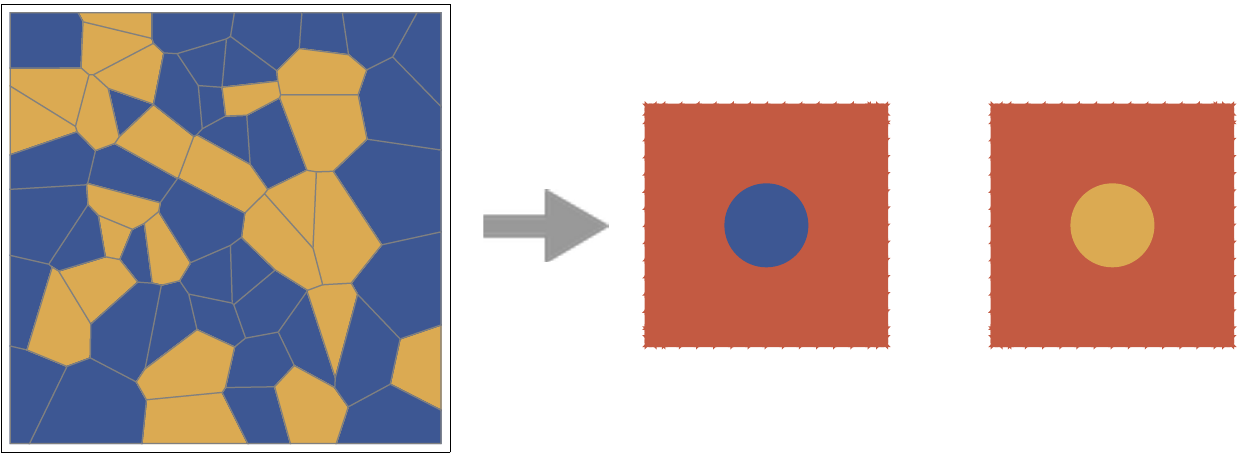}
\caption{A pictorial representation of Bruggeman's formalism.
On the left, the real system, which is composed by 
a mixture of two components characterized 
by their electrical conductivity.
On the right, the model, 
in which a single region (of either components) is assumed 
to be a sphere embedded in a single medium of uniform conductivity.
As an external field is turned on, the presence of the sphere causes a certain polarization.
By imposing that the average polarization due to all regions cancels out,
one deduces the relation Eq. \eqref{brugge}, 
which allows to determine the conductivity of the medium $\sigma_e$ in a self-consistent manner
\cite{landauer1978}.
}\label{figEMA}
\end{figure}

In the study of RS devices, electrical transport properties 
of inhomogeneous media are generally studied within the framework 
of percolation theory \cite{lee2015},
a statistical approach to the study of the formation of a percolation network 
\cite{stauffer1994,sahimi1994}.
RS indeed typically emerges as a consequence 
of an insulator-to-conductor transition of a composite material, 
occurring when the concentration of the conducting component crosses 
the so-called ``percolation threshold''.
Far from that threshold, as the case we are interested in,
the conductivity of an inhomogeneous medium can also be accurately 
estimated by means of an EMA
\cite{landauer1978,bergman1992,choy2015}.
Within such an approach, the inhomogeneous medium is 
effectively treated as a homogeneous one whose properties are calculated 
in a mean-field approximation. For regimes well above the percolation threshold, 
such simplification allows to accurately describe very complex systems 
with simple analytical formulas \cite{kirkpatrick1973,sahimi1994,luck1991}.
Among all possible realizations of the approach,
we adopt here Bruggeman's one \cite{bruggeman1935,landauer1978}, 
which allows to deal with a mixture of several components
treating them all on the same footing (see Fig. \ref{figEMA}).
Such an approach offers a natural framework for dealing 
with the different inhomogeneities enumerated in the previous section,
as we shall explain.

Given an inhomogeneous system made of a certain number
of homogeneous components, the conductivity of the system
is estimated via the ``effective conductivity'' $\sigma_e$
calculated as the positive root of the equation
\be\label{brugge}
\sum_i \Phi_i \frac{\sigma_i-\sigma_e}{\sigma_i+2\sigma_e}=0,
\ee
where $i$ runs over all the components, 
$\Phi_i$ is the relative volume fraction occupied by the component $i$,
for which $\sum_i \Phi_i=1$,
and $\sigma_i$ is the corresponding conductivity.
Such a formula concerns an arbitrary number
of conducting components mixed in the inhomogenous system.
An insulating component can be also considered 
by simply using a vanishing conductivity.
Indeed, a first, crude estimate the resistivity of a ng-film
can be obtained by considering the film as a mixture of 
bulk gold, for which $\sigma_{\rm Au}=0.041$ n$\Omega^{-1}$ m$^{-1}$ 
at room temperature \cite{serway1998},
and insulating vacuum, $\sigma_V=0$. 
Assuming that the nanoparticle density in the film
is close to that of a set of randomly packed spheres,
we consider $\Phi_{\rm Au}\approx0.63$ \cite{kishore1992},
and hence $\Phi_V=1-\Phi_{\rm Au}\approx 0.37$,
in Eq. \eqref{brugge} to estimate the resistivity of a ng-film to be 
$\rho\approx 55$ m n$\Omega$, which is twice as much that of bulk gold.
A more accurate estimate can be obtained if we improve the description
of the gold component. For instance, we can include the effects of grain boundaries
by taking its conductivity to be that 
of a polycrystalline film with similar grain size.
Using $\sigma_{\rm Au}\approx 0.01$ n$\Omega^{-1}$ m$^{-1}$,
which is the conductivity of a film with grains of approximately 10 nm
(see Ref. \onlinecite{henriquez2013} table 1, sample S1),
the resistivity of our ng-film is estimated to be
$\rho\approx 225$ m n$\Omega$.
This value, which is about ten times that of a crystalline film,
is comparable with the values reported by Mirigliano \textit{et al.}
of  $\rho \sim 100\div 1000$ m n$\Omega$ \cite{mirigliano2019},
which suggests that our approach is sound.

The flexibility of the EMA approach, in fact, allows us to be even more precise.
Suppose that within the highly inhomogeneous metal component
we can identify regions of relatively uniform conductivity,
which we say belong to the same ``phase'':
Eq. \eqref{brugge} can be accordingly reinterpreted 
as referring not to a mixture of just two components (vacuum and metal),
but to a mixture of different \emph{phases}, 
which can also belong to the same component (in our case, the metal one).
The classification of the various inhomogeneities of a ng-film
exposed in the previous section helps us in identifying phases in our systems.
For instance, if we believe that amorphous layers at the interfaces between
nanoparticles account for an important contribution to the total resistivity,
then one can single that contribution out by introducing \emph{three} phases:
a vacuum, an amorphous and a polycrystalline phase.
If, on the other hand, one believes that nanoparticle interfaces 
do not contribute more than simple grain boundaries,
but the size of nanoparticles is more important, 
one can introduce different phases to describe nanoparticles of different size.
For instance, in Ref. \onlinecite{mirigliano2019} it is reported that
two distinct nanoparticle populations characterized by a different radius (0.7 and 4.4 nm)
are clearly discernible, at least in the initial part of the deposition.
One can therefore introduce a phase for the population of smaller nanoparticles
and one for the larger ones.
A resistivity of about $\rho \sim 1000$ m n$\Omega$
is obtained, for instance, by considering four phases:
one for the vacuum, for which $\sigma_0=0$ and $\Phi_0=0.35$,
a first metallic phase, which might represent large nanoparticles,
with $\sigma_1=0.005$ n$\Omega^{-1}$ m$^{-1}$ and $\Phi_1=0.26$,
a second, less conducting metallic phase, representing small nanoparticles,
with $\sigma_2=0.001$ n$\Omega^{-1}$ m$^{-1}$ and $\Phi_2=0.26$,
and a third, even less conducting metallic phase,
representing amorphous layers between the nanoparticles,
with $\sigma_3=0.0005$ n$\Omega^{-1}$ m$^{-1}$ and $\Phi_3=0.13$.
The precise values for these quantities must be inferred either 
from further experimental measurements or atomistic simulations. 
For actual samples, they can presumably span a wide range of values. 
To get an idea of how much the resistivity estimate varies in different situations, 
in Table \ref{tabFIRSTESTIMATES} we report estimates for a variety of situations.
\begin{table}[h]
\caption{Resistivity estimates ($\rho$) for static four-phase ng-film,  
for different values of phase concentrations ($\Phi_i$) and relative conductivities ($\sigma_i/\sigma_{\rm Au}$, with $\sigma_{\rm Au}=0.041$ n$\Omega^{-1}$ m$^{-1}$),
keeping fixed the vacuum concentration and conductivity
to $35\%$ and $0$, respectively.}\label{tabFIRSTESTIMATES}
\begin{tabular}{cccccc|c}
\hline \hline
$\sigma_1/\sigma_{\rm Au}$ & $\sigma_2/\sigma_{\rm Au}$ & $\sigma_3/\sigma_{\rm Au}$ & $\Phi_1$ & $\Phi_2$ & $\Phi_3$ & $\rho$  \\ \hline
 0.122 & 0.0244 & 0.0122 & 26\%  & 26\%    & 13\%  & 1173 m n$\Omega$   \\
 0.122 & 0.0244 & 0.0122 & 32.5\%  & 32.5\%    & 0\%  &  950 m n$\Omega$   \\
 0.122 & 0.0244 & 0.0122 & 52\%  & 13\%    & 0\%  & 566  m n$\Omega$   \\
 0.122 & 0.0244 & 0.0122 & 13\%  & 52\%    & 0\%  & 1581  m n$\Omega$   \\
 1.000 & 0.0244 & 0.0122 & 26\%  & 26\%    & 13\%  & 509  m n$\Omega$   \\
 0.488 & 0.244 & 0.0122 & 26\%  & 26\%    & 13\%  & 180  m n$\Omega$   \\
\hline \hline
\end{tabular}
\end{table}


\section{Modeling dynamical processes}\label{sDyn}

\begin{figure}[t]
\centering
\includegraphics[width=0.5\textwidth]{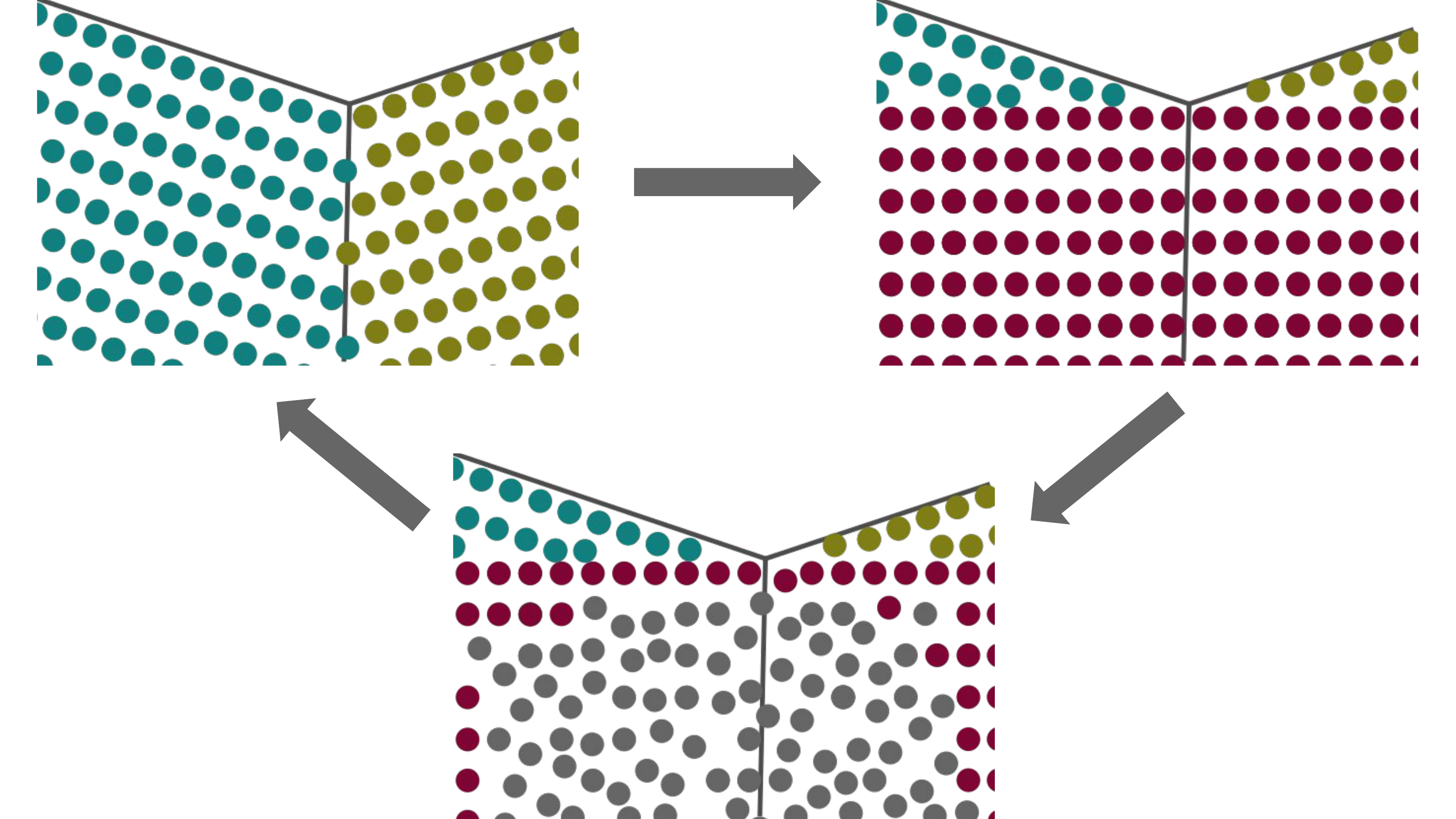}
\caption{
A sketch of the hypothetical microscopic mechanism proposed
in Ref. \onlinecite{mirigliano2020}
to explain RS in ng-films.
An active spot is identified with 
the interface between two nanoparticles (top left).
Such a system makes a transition to a higher conductivity state
via defect migration (top right);
from that, a lower conducting state is reached
because of local melting (bottom);
finally, the original interface is rebuilt (top left)
by action of the original crystals in the nanoparticles
that do not participate to the transition (inactive spots,
which surround the active ones).
}\label{figCYCLE}
\end{figure}

The discussion so far assumed that the system is in a static condition,
characterized by a ohmic response to an applied direct current.
In experiments, for a sufficiently strong applied current,
ohmic regimes are interrupted by RS phenomena,
which must reflect some sort of microscopical structural change of the system.
Due to the high degree of inhomogeneity of ng-films,
it is indeed reasonable to expect the local values of resistivity 
to have significant variations in time.
Defect migration, nanoparticle coalescence, 
and the melting at different temperatures of nanoparticles of different size  \cite{morris2008}
are all mechanisms that can contribute to various degrees to those variations.
In Ref. \onlinecite{mirigliano2020}, a cyclic mechanism
driven by the local flow of current,
here recalled in Fig. \ref{figCYCLE}, was proposed.
Lacking of direct evidence and the support of a quantitative analysis,
such a cycle is for now only hypothetical
and in the following we shall discuss the modeling 
of a rather more general class of dynamical mechanisms.
Nevertheless the cycle of Fig. \ref{figCYCLE}
will be used throughout our discussion
to show how our idea of modeling is carried out in practice.

For concreteness, we distinguish within the film regions
that undergo some structural change, ``active spots'', 
and regions that do not, ``inactive'' ones. 
In the case of Ref. \onlinecite{mirigliano2020},
an active spot is the boundary between two adjacent nanoparticles.
The scale of the actual mechanism(s) may be larger,
and therefore an active spot can in fact
comprehend various nanoparticles.
Starting from a given initial atomic configuration,
an active spot experiences over time a rearrangement of its atoms,
as effect of temperature variations
due to the current flowing through it.
A rearrangement can either lead to a configuration with 
\emph{higher} conductivity,
if, for example, two nanoparticles coalesce or a defect migrates,
or to a \emph{lower} conductivity,
as effect, for instance, of local melting.
In time, several rearrangement may occur, and it is not to exclude 
the possibility that
an active spot may be later found in its initial configuration,
as seen in Fig. \ref{figCYCLE}.

The presence of an onset time of a few seconds 
for the resistance jumps to start 
suggests that in this transient regime the spots are not active yet.
We argue that a current is needed for the Joule heating to provide
the missing energy to activate the spots.
More specifically, we conceptualize the process sequence as follows:
the activity of a spot is initiated
as soon as the temperature of the spot 
exceeds a critical temperature $\bar{T}$: 
when this situation is reached,
the spot has sufficient energy
to overcome the activation barrier 
that prevented the first rearrangement to occur.
In fact, it is reasonable to expect that different kinds
of structural rearrangement which an active spot
may go through are affected in one way or another 
(i.e. activated or deactivated, at least to some degree)
by the value of temperature at the spot.
We therefore suppose that a few critical 
temperatures $\bar{T}_1, \bar{T}_2,...$, 
which mark different energy thresholds
that allow or forbid the relevant rearrangements, exist.

\begin{figure*}[ht]
\centering
\includegraphics[width=0.98\textwidth]{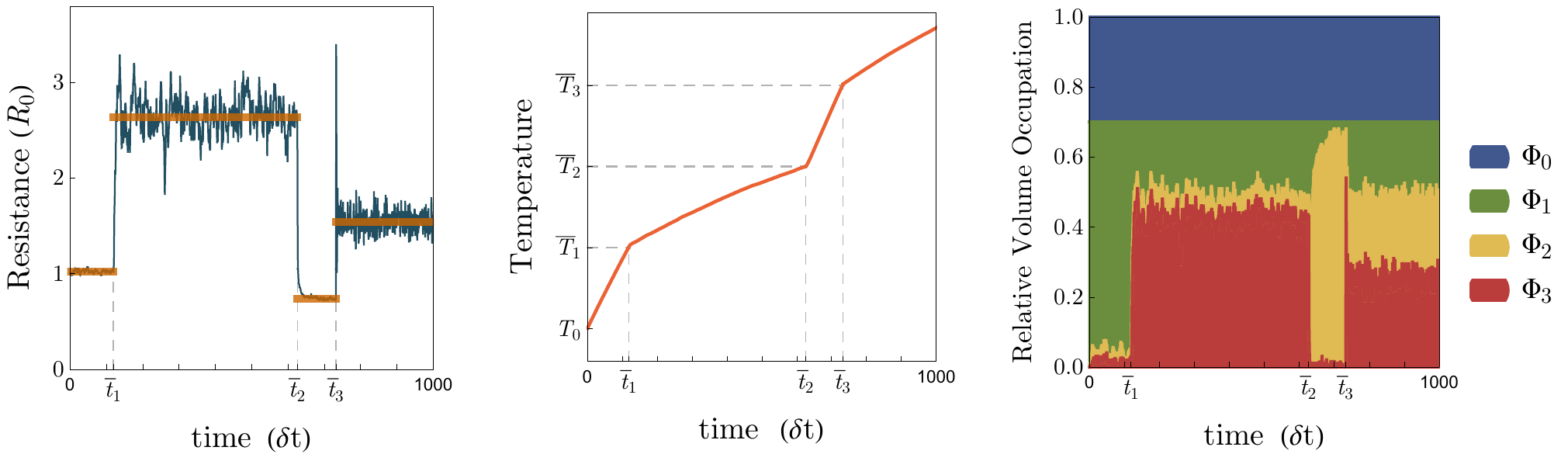}
\caption{Relative resistance (left),
temperature (center),
and relative volume occupied by the different phases (right)
in a paradigmatic simulation of a ng-film 
with thermally regulated microscopic processes.
A simple statistical analysis allows to calculate the average values
of resistance (left, orange straight lines)
without the need to run the simulation.
}\label{figNEAT}
\end{figure*}

The actual evolution of an active spot is therefore determined by
current-related fluctuations of the local temperature. 
In order to describe the overall dynamical evolution,
we adopt a stochastic approach. More specifically,
we consider a discrete time variable $t_a$,
characterized by a step $\delta t=t_{a+1}-t_a$ 
that can be tuned to match the experimental sampling frequency
(e.g. $\delta t=$  0.1 s in Ref. \onlinecite{mirigliano2019}).
The dynamics of the system is determined by a rule
that dictates how each single active spot changes its state
from one timestep to the next.
We assume a Markovian evolution,
according to which at each timestep an active spot
has a given probability of changing its state
that only depends on its current state and the ng-film temperature.
A complete characterization of the system is therefore
achieved by specifying the corresponding transition probability matrix $P$.
If, for instance, we assume a thermally regulated three-state cycle
as in Ref. \onlinecite{mirigliano2020},
the transition probability matrix can be written as
\bea\label{tp} 
P(T)&=&
\left(
\begin{matrix}
\mathcal{P}_{1\to1} & \mathcal{P}_{1\to2} & \mathcal{P}_{1\to3}\\
\mathcal{P}_{2\to1} & \mathcal{P}_{2\to2} & \mathcal{P}_{2\to3}\\
\mathcal{P}_{3\to1} & \mathcal{P}_{3\to2} & \mathcal{P}_{3\to3}\\
\end{matrix}\right)\\
&=&\left(\begin{matrix}
1-p_{12}(T) & p_{12}(T) & 0\\
0 & 1-p_{23}(T) & p_{23}(T)\\
p_{31}(T) & 0 & 1-p_{31}(T)\\
\end{matrix}\right)
\eea
where the only three degrees of freedom of the matrix
are parametrized by functions of the ng-film temperature $p_{nm}(T)$
for which $0\leq p_{nm}(T)\leq 1$ and are taken to suddenly change 
around some critical temperatures $\bar{T}_1$, $\bar{T}_2$, ...
Such a parametrization takes into account the fact 
that transitions can only happen in one direction
(so, say, $\mathcal{P}_{1\to 2}>0$ but $\mathcal{P}_{1\to 3}=0$)
and ensures that total probability 
$\mathcal{P}_{n\to1}+\mathcal{P}_{n\to2}+\mathcal{P}_{n\to3}$ sums up to 1.
The functions $p_{nm}(T)$ are characteristic of the microscopic
mechanism one wants to model and their exact form can 
in principle be determined by means of
numerical simulations of a single active spot.
As they are function of the temperature, we must also specify how that quantity evolves with time.

Temperature variations of the ng-film happen for two reasons, namely
{\rm i}) Joule heating, for which
\be 
\Delta T\;m\;c=\Delta t\; V^2/R
\ee
$V$ being the applied voltage, $R$ the resistance of the ng-film,
$m$ its mass and $c$ its specific heat,
and {\rm ii}) heat dissipation due to contact with the environment,
for which
\be
\Delta T=-\frac{\Delta t}{\tau} \big[T(t)-T_{env}\big]
\ee
with $\tau$ a characteristic time constant
that describes how fast the film heat is transferred to the substrate
and $T_{env}$ the environment temperature.
Combining the two contributions into a law for discrete time propagation we can write
\be\label{tdT}
T_{a+1}=T_a+\big[c_1 \sigma_e(t_a)-c_2(T_a-T_{env})\big]\delta t
\ee
where $c_1$ and $c_2$ are two suitable constants
and the film conductivity $\sigma(t_a)$ is calculated via 
a time-dependent version of Eq. \eqref{brugge}, namely
\be\label{tdbrugge}
\sum_{i} \Phi_i(t_a)\frac{\sigma_i-\sigma_e(t_a)}{
        \sigma_i+2 \sigma_e(t_a)}=0
\ee
where the index $i$ runs over all the phases identified in the system
while $a$ represents the time index.
In the case of the mechanism of Fig. \ref{figCYCLE}
we would introduce four phases: 
an insulating one for describing the voids,
a polycrystalline characterized by a certain grain size 
(corresponding to the top left state),
a second polycrystalline characterized by bigger grains (top right),
and an amorphous one (bottom).

If the temperature of the sample spans a wide range of values,
the conductivity of the single phases may vary
due to the increasing contribution of electron-phonon scattering events,
in which case we must consider 
\be\label{tdbrugge1}
\sum_{i} \Phi_i(t_a)\frac{\sigma_i(T_a)-\sigma_e(t_a)}{
        \sigma_i(T_a)+2 \sigma_e(t_a)}=0
\ee
instead of Eq. \eqref{tdbrugge}.
Around room temperature, the resistivity of gold, from simple bulk \cite{giancoli2005} 
to very complex systems like nanocrystalline films \cite{ederth2000},
increases linearly with the temperature and 
therefore we consider $\sigma_i(T_a)=\sigma_{i(0)}[1+\alpha_i (T_a-T_{env})]^{-1}$,
where $\alpha_i$ a positive parameter usually called Temperature Coefficient of Resistivity, TRC.
A signature of a nonvanishing TRC is a slight drift in an ohmic regime
towards higher values of resistivity,
which may be recognized in the resistance measurements of a ng-film
under the bias of 0.5 V in Ref. \onlinecite{mirigliano2019}.

Provided with numerical values for conductivity 
$\sigma_{i(0)}$ and the TRC $\alpha_i$ of the different phases, 
the parameters  $c_1$, $c_2$ of the temperature equation \eqref{tdT},
a specific behavior for the functions $p_{nm}(T)$,
the total number of spots and the phase initial densities,
it is possible to propagate the model and simulate 
the time evolution of a ng-film.
In Fig. \ref{figNEAT}
we present a specific realization of such a model,
whose parameters, reported in Table \ref{tabPAR} and \ref{tabPNMNEAT},
were chosen to give rise to the relevant features
of the RS behavior observed 
in the experiments of Ref. \onlinecite{mirigliano2019}.
We can indeed clearly distinguish a big spike
and jumps between ohmic regimes characterized by different duration, 
fluctuation amplitude and resistance value,
which can be higher as well as lower than the initial value.
The ohmic regimes are characterized by fluctuations
due to the occurrence of the structural rearrangements of 
some active spots at each time step.
The amplitude of the fluctuations reduces 
if the number of active spots is increased.
A comparison between the temperature and resistance evolution
shows that steps occur every time a critical temperature is crossed.
In other words, whenever the transition probabilities are stable,
only fluctuations occur, while jumps arise when transition probabilities
suddenly change. 
The duration of a ohmic regime is therefore linked to the time the 
sample takes to reach a critical temperature.
As in the experiments, a voltage threshold,
under which no jumps but only fluctuations occur, 
is also observed, as result of the fact that Joule heating,
which is proportional to $V^2$,
is not sufficient to take the ng-film to cross the lowest critical temperature.
Rising the voltage well above the threshold can significantly accelerate
the dynamical evolution, as critical temperatures are reached in a shorter time.

Such a simulation therefore teaches us that jumps of resistivity, 
such as those occurring at $t_1$, $t_2$ and $t_3$ in Fig. \ref{figNEAT},
are not caused by structural changes involving single active spots 
(e.g. the melting of a juction of the migration of a grain boundary).
Far from the percolation threshold it would take much more 
than a single boundary to melt to give rise to a macroscopic jump. 
In the simulated sample, such microscopic events happen very frequently 
and in fact they are at the origin of the ``noise'' one can see 
during the periods of stability of the resistance. 
What causes the jumps is the sudden prevalence of one phase over the others,
as shown in the last plot of Fig. \ref{figNEAT}. 
In other words, during the entire time of the measuring process, 
the relative concentration of the three phases depicted in Fig. \ref{figCYCLE}
is mostly stable, but sometimes it changes quite abruptly. 
This is a consequence of the fact that the transitions of 
an active spot from one state to another are (de)activated 
at certain temperatures and, each time the sample crosses a critical temperature, 
a certain state becomes dominant over the others.

A statistical analysis allows
to exactly calculate the values of resistivity spanned 
during the ohmic regimes starting from 
the transition probability matrix defined in Eq. \eqref{tp} 
and the values of conductivity of the single phases.
During such regimes,
the temperature of the ng-film is between 
two critical temperatures and the transition probability matrix
is constant, $P(T)= P$. We can then say  \cite{stewart2009}
that the average concentration of the different phases $\braket{\Phi}_i$ 
is given by $\braket{\Phi}_i = P^{\infty}_{i1}$ 
where $P^{\infty}$ is the limiting
transition matrix $P^{\infty}=P P P....$,
for which $P^{\infty}_{1m}=P^{\infty}_{2m}=P^{\infty}_{3m}$.
Provided with  $\braket{\Phi}_i$ for each ohmic regime,
one can use  Eq. \eqref{brugge} to calculate the effective conductivity,
and hence the resistance of the film, for the such regimes
without performing a propagation of the stochastic equations.
Such an analysis,
whose numerical outcome for this specific case 
is reported in Fig. \ref{figNEAT}
on top of the stochastic simulation,
is particularly important for cases in which
the conditions of the simulation, or, for that matter, of an experiment,
cannot ensure that the entire landscape 
of allowed values is spanned.


\begin{figure}[ht]
\centering
\includegraphics[width=0.47\textwidth]{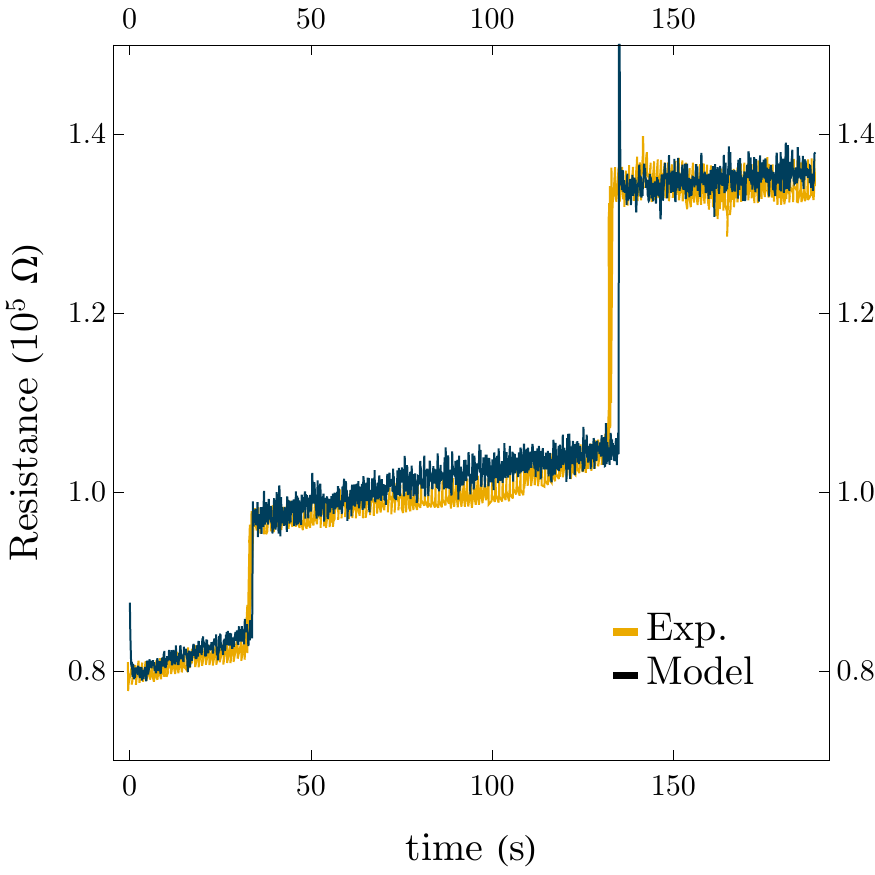}
\caption{
Resistance of a simulated ng-film
with microscopic parameters tailored to give rise
to the RS reported 
in Ref. \onlinecite{mirigliano2019}, Fig. 4(c).
}\label{figMY4C}
\end{figure}




Finally, to prove that the approach can quantitatively,
and not only qualitatively, explain the experimental data,
in Fig. \ref{figMY4C} we show another realization of the model,
whose parameters were tuned to match the experimental results for the ng-film
with thickness 30 nm under the bias of 0.5 V, reported in 
Ref. \onlinecite{mirigliano2019}, Fig. 4(c).
In fact, since the same graph $R(t)$ can be obtained
with various choices of the parameters,
we assumed certain features of the spots
that one would expect from the specific mechanisms 
represented in Fig. \ref{figCYCLE}.
More specifically, we assumed
a certain ratio between the conductivity of the different phases
and an overall behavior of the functions $p_{nm}(T)$
characterized by two critical temperatures,
one corresponding to the defect migration
activation temperature and the other to 
the local melting temperature.
Given then the specifics of the experiment
(sample size and density, and frequency of the current measurements),
we tuned the two parameters of the conductivity of the first phase 
$\sigma_{1(0)}$ and $\alpha_1$,
which set a scale for the resistance,
the constant $c_2$, 
which determines the time scale of the problem,
and the number of active spots,
which sets the amplitude of the fluctuations
in the ohmic regimes, to match the corresponding
features of the experimental data 
(\textit{cf} Tables \ref{tabPAR} and \ref{tabPNMMY4C}).
Such a careful choice of the parameters
allows to reproduce the experimental data with 
a sufficient level of accuracy.
Such an accuracy, however, must not be mistaken 
with a proof that the modeled microscopic processes indeed occur 
and the values used for the parameters, although plausible, 
reflect their real values. 
The available information on the experimental setup is not enough 
to sufficiently restrict the space of parameters compatible 
with the experimental data. 
The same macroscopic resistance can be achieved with 
quite different choices of the parameters.
In fact, also the details of the modeled microscopic mechanism, 
including even its cyclicity, are not necessary ingredients 
to reproduce the observed data; 
what is important is the presence of thermally
(de)activated local structural rearrangements.
The level of accuracy of the simulation therefore only supports
the adequacy of the mathematical tools we here put together
for describing this form RS we are interested in.

\begin{table}[h]
\caption{List of the parameters, except transition probabilities, 
used to obtain Fig. \ref{figNEAT} and Fig. \ref{figMY4C}. 
While for the latter fully dimensional values were used,
for the former the parameters are expressed in terms 
of three quantities (one for the conductivity, one for the temperature and the timestep)
that set the scale of the problem.}\label{tabPAR}
\begin{tabular}{lll}
\hline \hline
                & Fig. \ref{figNEAT} & Fig. \ref{figMY4C} \\ \hline
length          &                                     & 1 mm                                \\
width           &                                     & 1 mm                                \\
thickness       &                                     & 30 nm                               \\
$\sigma_0$      &             0                       & 0                                   \\
$\sigma_{1(0)}$ &                                     & 1450 m$^{-1}$ $\Omega^{-1}$         \\
$\alpha_1$      &             0                       & 0.00013 K$^{-1}$                    \\
$\sigma_{2(0)}$ &           1.4 $\sigma_{1(0)}$       & 2175 m$^{-1}$ $\Omega^{-1}$         \\
$\alpha_2$      &             0                       & 0.00013 K$^{-1}$                    \\
$\sigma_{3(0)}$ &           0.2 $\sigma_{1(0)}$       & 725 m$^{-1}$ $\Omega^{-1}$          \\
$\alpha_3$      &             0                       & 0.00013 K$^{-1}$                    \\
Active spots    &             200                     & 2000                                \\
Inactive spots  &             0                       & 0                                   \\
$\Phi_0     $   &            0.3                      & 0.504916                            \\
$T_0$           &                                     & 300 K                               \\
$\bar{T}_1$     &            $(1+10^{-5})T_0$         & 710 K                               \\
$\bar{T}_2$     &            $(1+2\times10^{-5})T_0$  & 1392 K                              \\
$\bar{T}_3$     &            $(1+3\times 10^{-5})T_0$ &                                     \\
$T_{env}$       &            $T_0$                    & 300 K                               \\
$c_1$           &$(6\times 10^6)^{-1}T_0(\sigma_{1(0)}\delta t)^{-1}$& 0.0324 K m$^{3}\;\Omega$ s$^{-1}$\\
$c_2$           &      $10^{-3}\delta t^{-1}$         & 0.005 s$^{-1}$                      \\
$\delta t$      &                                     & 0.1 s \\
\hline \hline
\end{tabular}
\vspace{0.5cm}

\caption{Values of the functions that identify the
transition probability matrix Eq. \eqref{tp} for the realization 
of the model leading to Fig. \eqref{figNEAT}.}\label{tabPNMNEAT}
\begin{tabular*}{\columnwidth}{@{\extracolsep{\fill} } r|c|c|c|c}
\hline \hline
&$T<\bar{T}_1$&$\bar{T}_1<T<\bar{T}_2$&$\bar{T}_2<T<\bar{T}_3$&$\bar{T}_3<T$\\\hline
 $p_{12}$ & 0.01           &  0.2  & 0.1 & 0.8\\
 $p_{23}$ & 0.3           &  0.8  & 0.008 & 0.8\\
 $p_{31}$ & 0.4           &  0.1  & 0.7 & 0.7 \\
 \hline\hline
\end{tabular*}
\vspace{0.5cm}
\caption{Values of the functions that identify the
transition probability matrix Eq. \eqref{tp} for the realization 
of the model leading to Fig. \eqref{figMY4C}.}\label{tabPNMMY4C}
\begin{tabular*}{\columnwidth}{@{\extracolsep{\fill} } r|c|c|c}
\hline \hline
& $T<\bar{T}_1$ & $\bar{T}_1<T<\bar{T}_2$  &$\bar{T}_2<T$   \\ \hline
 $p_{12}$ & 0.2           &  0.9  & 0.2 \\
 $p_{23}$ & 0.1           &  0.3  & 0.8 \\
 $p_{31}$ & 0.7           &  0.7  & 0.3\\
 \hline \hline
\end{tabular*}
\end{table}

\section{Conclusions}\label{sCon}

Nanogranular gold films with thickness well beyond the percolation threshold
present an intriguing, unexpected dynamical response 
to external electrostatic potentials.
Understanding and, ultimately, harnessing such a phenomenon can
lead to interesting technological applications.
Their modeling, however, poses some theoretical challenges.

Even before introducing dynamical effects, 
estimating the resistivity of a nanogranular film
eludes the standard models used for crystal films 
which do not take into account the complex structure 
arising from the presence of nanoparticles.
In this work we suggest to approach the problem 
using Bruggeman's formalism for multicomponent media,
which belongs to the framework of the Effective Medium Approximations.
The approach was designed for macroscopic scales \cite{bergman1992}
and is expected to be accurate only
when a sufficiently large number of percolation paths
have been established \cite{kirkpatrick1973}
(a regime that for inhomogeneous systems is not always 
easy to establish by simply looking 
at the overall resistance \cite{falcon2017}).
Apart from those caveats,
which may be resolved by resorting to 
more sophisticated approximations \cite{grimaldi2014},
the method is typically rather accurate,
computationally very inexpensive and, 
most importantly for our case, quite flexible. 
We have indeed explained how it can be adapted
to describe not only the porosity,
but also other features of ng-films, such as
amorphous layers at the nanoparticle interfaces
and the presence of nanoparticles of different size,
complementing other models designed to account
for the remaining sources of RR.

Provided with such a tool it is possible to build dynamical models.
To give a concrete example of the flexibility and extent of the approach, 
we built a stochastic model connecting
the macroscopic film resistance to
a generic class of microscopic mechanisms,
characterized by local variations of resistivity
due to structural rearrangements.
Two specific realizations of the model,
with tailored choice of the parameters,
were shown to reproduce, 
qualitatively as well as quantitatively, 
the characteristic jumps of resistivity,
as well as other, less obvious, 
features of the experimental data.
Our simulations proved that, 
far from the percolation threshold,
in a regime where local structural rearrangements 
are not sufficient to explain macroscopic jumps of resistance,
RS can arise if those rearrangements 
are thermally (de)activated;
as the sample temperature rises as effect of Joule heating,
certain rearrangements become more favorable than others
and when such a transition happen an appreciable jump of resistance occurs.
Since local structural rearrangements
happening in nanogranular films
can be investigated by means of atomistic simulations
that only have to involve a few nanoparticles,
such a model can be considered as a starting point
for further theoretical investigations 
aimed at shading light upon the such mechanisms.


In summary, we presented a modeling based on EMA 
that can be used to study
the RS phenomena that seems to characterize 
nanogranular film also beyond the percolation threshold
and we showed how simple thermally regulated local structural changes
can explain the most relevant features 
of the experiments, 
suggesting a clear path for future investigations.

\begin{acknowledgments}
This work was fully funded by Fondazione CON IL SUD (Grant No: 2018-PDR-01004).
\end{acknowledgments}

\bibliography{main.bbl}

\end{document}